\documentclass[a4paper,11pt]{article}
\usepackage{pos}

\newcommand{\vect}[1]{\boldsymbol{#1}_{\perp}}
\newcommand{\kt}{\vect{k}}

\newcommand{\Pt}{\vect{P}}

\newcommand{\qt}{\vect{q}}

\newcommand{\bt}{\vect{b}}

\newcommand{\rt}{\vect{r}}

\newcommand{\der}{\mathrm{d}}

\newcommand{\Tr}{\mathrm{Tr}}

\title{Quark fracture function at small $x$ from the Color Glass Condensate}

\author*[a]{Paul Caucal}
\author[a,b]{Nolann Juet}
\author[c,d,e]{Farid Salazar}

\affiliation[a]{SUBATECH UMR 6457 (IMT Atlantique, Universit\'{e} de Nantes,
IN2P3/CNRS), 4 rue Alfred Kastler, 44307 Nantes, France}

\affiliation[b]{Institut Polytechnique de Paris, Route de Saclay, 91120 Palaiseau, France}

\affiliation[c]{Department of Physics, Temple University, Philadelphia, PA 19122 - 1801, USA}

 \affiliation[d]{RIKEN-BNL Research Center, Brookhaven National Laboratory, Upton, New York 11973, USA}

 \affiliation[e]{Physics Department, Brookhaven National Laboratory, Upton, New York 11973, USA}

\emailAdd{caucal@subatech.in2p3.fr}
\emailAdd{nolann.juet@polytechnique.edu}
 \emailAdd{farid.salazar@temple.edu}

\abstract{
The target-fragmentation region in deeply inelastic electron-nucleus scattering provides insight into the partonic structure of the target, as hadrons measured in the final state retain information about the nonperturbative dynamics of spectator partons inside the nucleus. In QCD factorization theorems, this nonperturbative dynamics is encoded in objects known as fracture functions. In this paper, we study the extended quark fracture function in the limit where the struck (anti)quark carries a very small longitudinal momentum fraction, $x$.  We first present an elementary derivation within the Color Glass Condensate effective field theory of the quark fracture function in terms of the dipole operator at small $x$. At the parton level, the quark extended fracture function is sensitive to gluon saturation when the transverse momentum $P_\perp$ of the outgoing parton produced in the target-fragmentation region is smaller than the nuclear saturation scale $Q_s$, in which regime the $P_\perp$ distribution becomes approximately flat. We then show, through a numerical study that incorporates the convolution of this parton-level result with a collinear fragmentation function, that hadronization largely washes out these saturation effects, as the convolution predominantly probes partonic transverse momenta that are typically larger than $Q_s$.}

\FullConference{The 33rd International Workshop on Deep Inelastic Scattering and Related Subjects (DIS2026)\\
4 - 8 May 2026\\
Bologna, Italy\\}

\begin{document}
\maketitle

\section{Introduction}

Measuring gluon saturation, the emergent QCD phenomenon associated with the taming of the growth of the gluon density in nuclei and the unitarization of QCD cross sections at sufficiently high energies, is a central goal of modern QCD phenomenology~\cite{Morreale:2021pnn}. In particular, it is an important part of the scientific program of the future Electron-Ion Collider (EIC)~\cite{Accardi:2012qut}, which is expected to collide electrons with heavy nuclei for the first time around 2035.

Very recently, significant activity has been devoted to the study of the target fragmentation region (TFR) in Deep Inelastic electron-nucleus scattering in the small Bjorken-$x$ limit~\cite{Liu:2022wop,Liu:2023aqb,Chen:2024bpj,Caucal:2025qjg,Mantysaari:2026zte}. Although unconventional, since most DIS observables at leading order have so far been constructed at small $x$ from the outgoing quark-antiquark pair produced by the decay of the virtual photon in the current fragmentation region of the Breit frame, the target fragmentation region may offer new opportunities to probe gluon saturation through the measurement of jets and hadrons originating from spectator partons inside the incoming nucleus~\cite{Liu:2023aqb,Caucal:2025qjg}. In the TFR, for transverse momenta of the measured hadron much smaller than the virtuality of the process, the cross section factorizes into a hard part and a non-perturbative object known as an extended fracture function. Extended fracture functions $M^i_{A,h}(x,\xi_h,P_\perp)$ were first introduced in~\cite{Trentadue:1993ka,Grazzini:1997ih} and can be interpreted as the conditional probability to find a parton $i$ carrying a longitudinal momentum fraction $x$ in a nucleus $A$, given that a spectator parton hadronizes into a hadron $h$ with longitudinal momentum fraction $\xi_h$ and transverse momentum $P_\perp$. They are related to diffractive distributions~\cite{Collins:1997sr,Berera:1994xh} in that they share the same operator definition~\cite{Hautmann:1999ui}. The latter, however, are more restrictive quantities, since the measured hadron is required to be the recoiling nucleon and rapidity gaps are imposed in diffractive events, neither of which is required here. In the small-$x$ limit, for sufficiently large saturation scale $Q_s$, the extended fracture functions can be computed within the Color Glass Condensate (CGC) effective field theory~\cite{Caucal:2025qjg}. 

The goal of this paper is to present a proof-of-concept study of the quark extended fracture function in the presence of gluon saturation, focusing on its measurement in semi-inclusive DIS through the tagging of a hadron in the TFR. Compared with the results presented in~\cite{Caucal:2025qjg}, we consider hadronic rather than jet final states and also include the $\xi_h$ dependence of the fracture function. Although jets are under better theoretical control, they may be more difficult to measure in the low-$P_\perp\lesssim Q_s$ region where saturation effects arise. It is therefore important to assess whether a hadronic measurement retains sensitivity to saturation effects. Our main conclusion is that hadronization washes out the saturation transition present in the parton-level or quark jet fracture function.

\section{A simple derivation of the quark extended fracture functions in the CGC}

 We aim at computing the cross-section for producing a hadron in the TFR in $\gamma^*A$ scattering,
 \begin{align}
    \frac{\der\sigma^{\gamma^*_TA\to h+X}}{\der^2\Pt\der \xi_h}\,,\label{eq:xsection-def}
\end{align}
 where, for simplicity, we restrict ourselves to the transversely polarised photon case. We work in the Breit or dipole frame, where both the virtual photon and incoming nucleus have zero transverse momentum. In light-cone coordinates, the photon four-vector is $q^\mu=(q^+,-Q^2/(2q^+),\boldsymbol{0}_\perp)$ and the nucleon from the nucleus four vector is $P^\mu=(0,P^-,\boldsymbol{0}_\perp)$ (we neglect target mass corrections). One measures the transverse momentum $\Pt$ and the minus longitudinal momentum fraction $\xi_h=k^-_h/P^-$ of the hadron with respect to the target. The limit we are interested in is both the high energy limit $W^2\gg Q^2$ --- with $W^2=(q+P)^2$ and $Q^2=-q^2$ ---, and small transverse momentum $Q^2\gg P_\perp^2$ where one expects factorization of the cross-section into a hard and non-perturbative parts.

\begin{figure}
    \centering
    \includegraphics[width=0.65\linewidth]{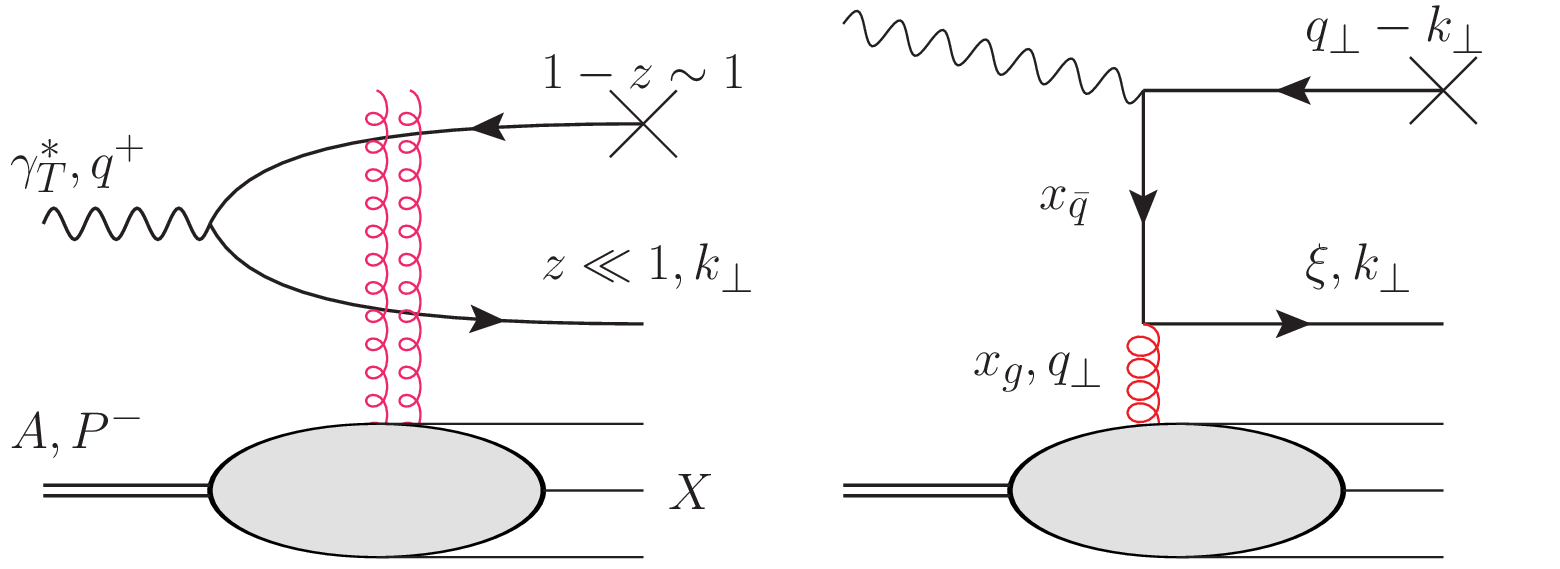}
    \caption{Projectile picture in the CGC (left) and target picture (right) graphs for SIDIS in the target fragmentation region. The cross on the final state antiquark means that the particle is not measured.}
    \label{fig:graph}
\end{figure}

Our starting point to compute eq.\,\eqref{eq:xsection-def} is the leading-order expression for the $\gamma^*_T+A\to q+X$ cross-section in the CGC obtained from squaring the amplitude associated with the left graph in Fig.\,\ref{fig:graph}. Written in momentum space, the cross-section differential w.r.t.~the transverse momentum $k_\perp$ of the quark and its plus longitudinal momentum fraction $z=k^+/q^+$ reads (see e.g.~\cite{Marquet:2009ca,Caucal:2024vbv})
\begin{align}
    \frac{\der \sigma^{\gamma_{\rm T}^{\star}A\to q+X}}{  \der^2 \kt\der z} &=\frac{\alpha_{\rm em}e_q^2N_c}{2\pi^2}(z^2+(1-z)^2) \int\der^2\bt\int\der^2\qt \ \mathcal{D}_F\left(x_g,\bt,\qt\right) \nonumber\\
    &\times \left\{\frac{\kt^2}{[\kt^2+\bar Q^2]^2}+\frac{(\kt-\qt)^2}{\left[(\qt-\kt)^2+\bar Q^2\right]^2}-\frac{2\kt\cdot(\kt-\qt)}{[(\kt-\qt)^2+\bar Q^2][\kt^2+\bar Q^2]}\right\}\,,\label{eq:LO-SIDIS-full}
\end{align}
with the dipole operator defined in terms of light-like fundamental Wilson lines $V$ as
\begin{align}
\mathcal{D}_F(x,\bt,\qt)\equiv\int\frac{\der^2\rt}{(2\pi)^2}e^{-i\qt\cdot\rt}\frac{1}{N_c}\left\langle \Tr(V(\bt+\rt/2)V^\dagger(\bt-\rt/2))\right\rangle_x\,,
\end{align}
and the effective virtuality $\bar Q^2\equiv z(1-z)Q^2$. Eq.\,\eqref{eq:LO-SIDIS-full}  is valid at any $z$. In the limit $Q^2\gg k_\perp^2$, the cross-section is controlled by the two endpoints $z\sim k_\perp^2/Q^2$ and $1-z\sim k_\perp^2/Q^2$ --- it is the so called "aligned-jet" configurations --- in order for $\bar Q^2$ to be of the same order as $k_\perp^2$ and therefore not to induce an additional $1/Q^2$ suppression. However, for a quark produced in the TFR, only the endpoint $z\sim k_\perp^2/Q^2$ contributes since the other endpoint corresponds to a quark almost aligned with the virtual photon direction, and therefore produced in the current fragmentation region.

In this regime, it is more convenient to deal with the minus longitudinal momentum fraction of the measured quark $\xi=k^-/P^-$. Using the onshellness of the quark, we find that
\begin{align}
    z=\frac{x_{\rm Bj}k_\perp^2}{\xi Q^2}\,.
\end{align}
Then, the next step is to simplify eq.\,\eqref{eq:LO-SIDIS-full} for $z\ll 1$ and to rewrite it in terms of $\xi$. As a result, we obtain
\begin{align}
   \frac{\der \sigma^{\gamma_{\rm T}^{\star}A\to q+X}}{  \der^2 \kt\der\xi }&=\frac{\alpha_{\rm em}e_q^2N_c}{2\pi^2Q^2} \int\der^2\bt\int\der^2\qt \ \mathcal{D}_F\left(x_{\rm Bj}+\xi,\bt,\qt\right)\frac{x_{\rm Bj}}{(x_{\rm Bj}+\xi)^2}\nonumber\\
    &\times \left[1+\frac{(x_{\rm Bj}+\xi)^2\kt^2(\kt-\qt)^2}{[\xi(\kt-\qt)^2+x_{\rm Bj}\kt^2]^2}-\frac{2(\xi+x_{\rm Bj})[\kt\cdot(\kt-\qt)]}{\xi(\kt-\qt)^2+x_{\rm Bj}\kt^2}\right]\,,\\
    &=\frac{4\pi^2\alpha_{\rm em}e_q^2}{Q^2}\times x_{\rm Bj}M^{\bar q}_{A,q}(x_{\rm Bj},\xi,\kt)\,,\label{eq:LO-SIDIS-TFR}
\end{align}
where the parton-level quark extended fracture function is defined as
\begin{align}
   M^{\bar q}_{A,q}(x,\xi,\kt)\equiv &\frac{N_c}{8\pi^4} \frac{1}{(x+\xi)^2} \int\der^2\bt\int\der^2\qt \ \mathcal{D}_F\left(x+\xi,\bt,\qt\right) \nonumber\\
    &\times \left\{1+\frac{(x+\xi)^2\kt^2(\kt-\qt)^2}{[\xi(\kt-\qt)^2+x\kt^2]^2}-\frac{2(\xi+x)[\kt\cdot(\kt-\qt)]}{[\xi(\kt-\qt)^2+x\kt^2]}\right\}\,.\label{eq:parton-level-qFrF}
\end{align}
To get eq.\,\eqref{eq:LO-SIDIS-TFR} we have stripped off the standard SIDIS leading twist hard factor for the $\gamma^*_T+\bar q\to \bar q$ subprocess in order to recognize in the $k_\perp$ dependence of the product the quark extended fracture function~\cite{Anselmino:2011ss}. Note that minus longitudinal momentum conservation for the hard subprocess $\gamma_T^*+\bar q\to \bar q$  leads to the kinematic constraint
\begin{align}
    q^-+x_{\bar q}P^-&=\frac{k_{\bar q\perp}^2}{(1-z)q^+}\simeq \frac{k_{\bar q\perp}^2}{q^+}\Rightarrow x_{\bar q}=x_{\rm Bj}\left[1+k_{\bar q\perp}^2/Q^2\right]\simeq x_{\rm Bj}\,,
\end{align}
where $x_{\bar q}$ is the minus longitudinal momentum fraction of the struck antiquark in the $t$-channel (see Fig.\,\ref{fig:graph}-right) and $k_{\bar q\perp}\simeq |\qt-\kt|\sim k_\perp$ its transverse momentum. On the other hand, minus longitudinal momentum for the process $g\to q\bar q$ gives $x_g=x_{\bar q}+\xi=x_{\rm Bj}+\xi$.
This explains why the argument of the dipole operator is set to $x+\xi$ in eq.\,\eqref{eq:parton-level-qFrF}. This equation has its domain of validity: first, one should have $x_g=x+\xi\ll 1$. Since $x\ll 1$, one must have $\xi\ll 1$ as well. However, we have assumed $z\ll 1$, and typically $z\sim k_\perp^2/Q^2$. This means that $\xi$ cannot be much smaller than $x$. So the above result is valid for $x \ll 1$ and $xk_\perp^2/Q^2\ll \xi\ll 1$.

It is easy to check that if one integrates $\xi$ between 0 and $1-x$ in this expression (neglecting the mild $\xi$-dependence inside $\mathcal{D}_F$) and expands for $x\ll 1$, one recovers the expression for $M^{\bar q}_{A,q}(x,\kt)$ from~\cite{Caucal:2025qjg}. Another cross-check is the dilute limit $k_\perp\gg Q_s$, where our result should match collinear factorisation~\cite{Chen:2021vby}. Indeed,
\begin{align}
      M^{\bar q}_{A,q}(x,\xi,\kt)&\sim \frac{\alpha_s}{4\pi^2}\frac{x^2+\xi^2}{(x+\xi)^3}\frac{1}{k_\perp^2}g(x+\xi,k_\perp^2)\,,\label{eq:large-kt}
\end{align}
with the (bare) gluon distribution function defined in terms of the dipole gluon TMD as
\begin{align}
    xg(x,\mu^2)&=\frac{N_c}{4\pi^2\alpha_s}\int\der^2\bt\int\der^2\qt  \ \qt^2\mathcal{D}_F\left(x,\bt,\qt\right)\Theta(\mu^2-q_\perp^2)\,.
\end{align}

Last but not least, to go from the parton level to the genuine fracture function into hadron, we convolute the above result with a collinear fragmentation function, such that
\begin{align}
        \frac{\der\sigma^{\gamma^*_TA\to h+X}}{\der^2\Pt\der \xi_h}=\frac{4\pi^2\alpha_{\rm em}e_q^2}{Q^2}\times  \left.xM^{\bar q}_{A,h}(x,\xi_h,\Pt)\right|_{x=x_{\rm Bj}}\,,
\end{align}
with
\begin{align}
    M^{\bar q}_{A,h}(x,\xi_h,\Pt)&=\int_{\xi_h/(1-x)}^1\frac{\der\zeta}{\zeta^3} \ D_{h/q}(\zeta,\mu_F^2) M^{\bar q}_{A,q}(x,\xi_h/\zeta,\Pt/\zeta)\,.\label{eq:qFrF-final}
\end{align}
This is the main result of this paper: it provides an explicit functional form of the quark extended fracture function in terms of a collinear quark fragmentation function and the dipole operator at small $x$. The factorization scale $\mu_F$ in the fragmentation function is set to $\mu_F=c k_\perp$ with $c$ varied between $1/2$ and $2$.

\section{Numerical study }

We now present some numerical results for the $P_\perp$ dependence of the quark to hadron extended fracture function resulting from eq.\,\eqref{eq:qFrF-final}. As noticed in~\cite{Caucal:2025qjg}, the quark-jet extended fracture function integrated over $\xi$ is sensitive to gluon saturation for $k_\perp\lesssim Q_s$. Our goal here is to check whether this remains the case for a hadronic measurement and for the $\xi_h$ differential case. We shall use $x_{\rm Bj}=0.003$ and $\xi_h=0.01$ such that $x_g=0.013$. 
These non too small values of $x_{\rm Bj}$ and $x_g$ should be realistic in view of the future EIC. We shall assume a homogeneous target such that $\int\der^2\bt=S_\perp$ where $S_\perp$ is the transverse area of the target. The dipole is evaluated using the GBW model~\cite{Golec-Biernat:1998zce} 
\begin{align}
\mathcal{D}_F(x,\bt,\qt)&=\frac{1}{\pi Q_s^2(x)}\exp\left[-\frac{\qt^2}{Q_s^2(x)}\right]\,,
\end{align}
where $Q_{s}^2(x)=A^{1/3}(x_0/x)^\lambda Q_0^2$ --- with $x_0=2.24\times 10^{-4}$, $\lambda=0.27$ and $Q_0=1$ GeV~\cite{Golec-Biernat:2017lfv} --- is the nucleus saturation momentum with atomic mass number $A$. The $x$ dependence of $Q_s$ approximates the high-energy Balitsky-Kovchegov (BK) evolution~\cite{Balitsky:1995ub,Kovchegov:1999yj} of the dipole, which effectively leads to an increase of $Q_{s}$ as $x$ decreases. For the fragmentation function, we consider two sets: JAM19~\cite{Sato:2019yez} and NNFF10~\cite{Bertone:2017tyb}, both embedded in the LHAPDF framework~\cite{Buckley:2014ana}.

\begin{figure}
    \centering
    \includegraphics[width=0.49\linewidth,page=1]{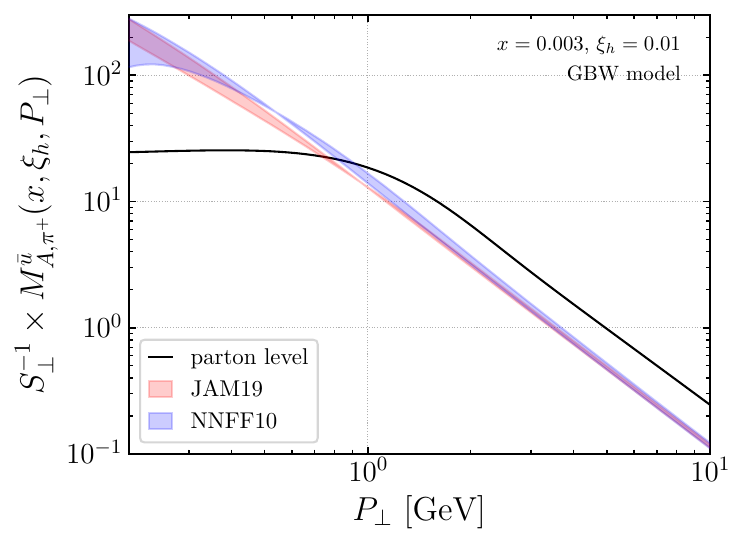}\hfill
    \includegraphics[width=0.48\linewidth,page=2]{plot-hadron-FrF.pdf}
    \caption{(Left) Quark extended fracture function into $\pi^+$ as a function of $P_\perp$ at fixed $x$ and $\xi_h$ for two sets of fragmentation functions. (Right) Ratio of quark extended fracture functions between nucleus and proton targets, scaled by $A^{-1/3}$, for two measured hadron species. Bands are obtained by varying $\mu_F$ in $[k_\perp/2,2k_\perp]$.}
    \label{fig:hadron-FrF}
\end{figure}

Our results are displayed in Fig.\,\ref{fig:hadron-FrF}-left, for a struck up antiquarks and a $\pi^+$ measured in the TFR. 
The black line is the parton level result given by eq.\,\eqref{eq:parton-level-qFrF}. One notices that although the GBW model does not correctly reproduce the perturbative high-$q_\perp$ tail of the dipole gluon TMD, it is sufficient for the present purpose, which is to understand to what extent saturation effects in the quark extended fracture function are suppressed by hadronization effects. Indeed, the GBW model qualitatively reproduces the main feature of the parton-level quark fracture function: (i) the $1/k_\perp^2$ perturbative tail at large $k_\perp$ given by eq.\,\eqref{eq:large-kt}, (ii) the constant behaviour for $k_\perp\lesssim Q_s$ due to saturation in the number of sea quarks inside the nucleus. Both behaviors are clearly visible on the black line of Fig.\,\ref{fig:hadron-FrF}. The red and blue curves correspond to two different choices of fragmentation function, JAM19 and NNFF10 respectively, which qualitatively display the same behaviour:  
in both cases, one notices that the taming of the distribution disappears due to the convolution with the fragmentation region which typically rescales the transverse momentum of the saturation transition to a much smaller value, from $Q_s$ to $\sim\langle \zeta\rangle Q_s$, with $\langle \zeta\rangle\sim 0.2$ the average or typical hadron to quark momentum fraction encoded in the collinear fragmentation function. On the right plot of Fig.\,\ref{fig:hadron-FrF}, we show the nuclear modification factor of the quark extended fracture function for two hadrons species, $\pi^+$ and $K^+$. Hadronization washes out both the small $P_\perp\lesssim Q_s$ suppression and the Cronin peak, although the situation looks slightly more favourable for kaons thanks to a larger $\langle \zeta\rangle$.

\section{Outlook}

To sum up, these results favour jet over hadronic measurement in the TFR (if experimentally feasible for semi-hard $P_\perp$ at the EIC) or the use of nucleon energy correlators (NECs)~\cite{Liu:2023aqb}. In any case, additional studies are necessary before reaching firm conclusions on the interest of measuring  particles in the TFR for saturation phenomenology. First, the gluonic sector needs to be considered~\cite{Chen:2024brp,Caucal:2025qjg,Mantysaari:2026zte}. Moreover, quantitative results require higher-order corrections, including large logarithms resummed to all orders via renormalization group equations. A simple extension of the results presented here would be to use a BK evolved dipole describing the $x_g$ dependence $\mathcal{D}$. In addition,  $M^i_{A,h}$ should depend on a additional resolution scale encoding the logarithmic dependence on the hard scale $Q$ of the process. The demonstration of the evolution equation resumming logarithms of $Q/k_\perp$ requires the one-loop correction in the CGC to the $\gamma^*_T+A\to q+X$ process considered here, which has recently been computed in~\cite{Caucal:2024cdq,Altinoluk:2025dwd}. These results should pave the way towards a full NLO result for the extended fracture functions or NECs in the small $x$ limit.

\textit{Acknowledgements.} P.~C. is funded by the Agence Nationale de la Recherche under
grant ANR-25-CE31-5230 (TMD-SAT). N.~J's research internship was supported by the JANUS-IN2P3 program. F.S. is supported by the Laboratory Directed Research and Development of Brookhaven National Laboratory and RIKEN-BNL Research Center.

\bibliographystyle{apsrev4-1}
\bibliography{refs}



\end{document}